\documentclass{aa}  

\usepackage{graphicx}
\usepackage{placeins}
\usepackage[version=3]{mhchem}
\usepackage{tabularx}
\usepackage{xcolor}
\usepackage[normalem]{ulem}
\usepackage{hyperref}
\hypersetup{colorlinks=true,linkcolor=blue,citecolor=blue,filecolor=blue,urlcolor=blue}
 
\usepackage{txfonts}

\begin{document}

   \title{Robust ellipticity measurements of 29 Galactic globular clusters}

\author{Laurane Fréour\thanks{\email{laurane.freour@univie.ac.at}}\inst{1}
    \and Ellen Leitinger\inst{2,3}
    \and Elena Pancino\inst{3}
    \and Alice Zocchi\inst{1}
    \and Glenn van de Ven\inst{1}
    }

\institute{Department of Astrophysics, University of Vienna, T\"urkenschanzstrasse 17, A-1180 Vienna, Austria 
\and Dipartimento di Fisica e Astronomia, Universit{\'a} degli Studi di Bologna, Via Gobetti 93/2, I-40129 Bologna, Italy
\and INAF- Osservatorio Astrofisico di Arcetri, Largo Enrico Fermi 5, I-50125 Firenze, Italy
}

\date{Received 11/08/2025 / Accepted 23/02/2026}
\titlerunning{Robust ellipticity measurements of 29 Galactic GCs}
 
  \abstract
   {\textit{Context:} Globular clusters (GCs) exhibit varying degrees of flattening (ellipticity), which may provide insight into their internal dynamics and evolution histories. Commonly used methods to measure ellipticity, such as ellipse fitting of density contours and principal component analysis, often produce biased results, especially for clusters that are nearly round or have few observable stars.
   \newline
   \textit{Aims:}
   Using a combination of ground-based and space-based photometry, we investigate the shapes of 29 Galactic GCs. To that end, we test two commonly used methods: an ellipse fit to a kernel density profile and a principal component analysis. We find that both methods suffer from bias arising when the number of stars is small or the cluster is close to round.
   \newline
   \textit{Methods:} To solve this issue, we develop a robust method to measure the ellipticity of GCs, test it extensively on mock data, and apply it to the 29 Milky Way GCs in our sample. Using the $V/\sigma$ diagram used in the isotropic oblate rotator framework, we examine potential causes for the flattening, including rotation and velocity anisotropy.
   \newline
   \textit{Results:}
   Our analysis reveals that 55\% of the clusters in our sample have a flattening superior to 0.05. For ten clusters: NGC~104, NGC~1261, NGC~2808, NGC 3201, NGC 5286, NGC 5904, NGC 5986, NGC 6205, NGC 6341, and NGC 7078 we identify a very good agreement between the rotation angle and semi-minor axis of the ellipse, further corroborating the findings that rotation is the main driver of the ellipticity. The $V/\sigma$ diagram reveals that velocity anisotropy or tides could also be important in shaping the GCs.
\newline
   \textit{Conclusion:} The robust method developed provides reliable measurements of the ellipticity of GCs, emphasising the importance of taking into account the flattening in theoretical models and simulations. It also offers a promising way to investigate the shapes of multiple stellar populations within GCs, where only small samples are usually available. Finally, the $V/\sigma$ diagram appears to be a good tool to understand the mechanism shaping GCs.
   }


\maketitle


\section{Introduction} \label{sec:intro}

It is now clearly established that globular clusters (GCs) are not as simple objects as previously thought, in their shape, their dynamics, and in their internal stellar populations. 

The first evidence of GCs differing from spherically symmetric objects was reported by \cite{1917Pease}. Using photographs taken by the 60-inch reflector at Mount Wilson, they manually counted the number of stars in different sectors. They concluded that several clusters, among them NGC~6205, had an elongated shape and were elliptical rather than spherical. Since then, many studies have focused on quantifying the ellipticity of GCs \citep{1982Frenk,1983Geyer,1987White,Chen_2010}, unravelling a flattening in a significant number of clusters. 

These findings indicate the importance of understanding the origin of the deviation from spherical symmetry of GCs. Determining whether this deviation arises from internal dynamics, such as rotation or velocity anisotropy, or from external tidal effects from the Milky Way (MW), allows us to uncover how GCs form, evolve, and interact with their environment.

Several studies have suggested rotation as the main driver of the shape of GCs. 
For example, the flattening of NGC~5139 \citep{1985Meylan,2003Pancino}, NGC~104 \citep{1984Mayor,1985Meylan,2017Bellini}, NGC~7089 \citep{1986Pryor}, and NGC~5904 \citep{2018Lanzoni} has been attributed mainly to rotation.
Using image processing of Palomar and SRC Sky Survey material, \citet{1987White}, hereafter WS87, studied the flattening of 100 Galactic GCs. They found that 32\% of their sample showed an axial ratio inferior to 0.9. By comparing the orientation of the semi-major axis of the GCs with the galactic center, they concluded that there was no evidence of tidal effects from the galaxy shaping GCs and that rotation or velocity anisotropies were more likely to contribute to the shape of the cluster. 
\citet{2014Fabricius} found a tight correlation between the ellipticity of the cluster and its central rotation in 11 MW GCs. Using 3D kinematics data, \citet{2024Dalessandro} found a correlation between their newly introduced $\alpha$ parameter, probing the relative strength of the rotation signal over the disordered motion, and their best-fit ellipticity values, supporting the conclusion that flattening is linked to rotation.
\cite{2018Kamann} also noticed a trend with more elliptical clusters corresponding to higher rotation. They highlighted that other mechanisms, such as velocity anisotropy or tidal effects, could blur this correlation. 
This last point has also been emphasised by \citet{2019Sollima} as a potential reason for the absence of correlation between ellipticity and rotation. Using ellipticities from \citet{Chen_2010}, hereafter CC10, and rotations derived from Gaia DR2 on 62 MW GCs, they did not find any correlation between these two parameters, nor between the isophotal minor axis and rotation axis. Such a conclusion is also supported by the results from \citet{2018Bianchini}. They computed the $V/\sigma$ parameter as the ratio between the peak of the rotation that they measured and the central velocity dispersion from \citet{baumgardt_Hilker_2018}, and did not find any significant correlation with ellipticity, using the values from WS87.
{On the contrary, \citet{2024Cruz_reyes}, hereafter CR24, noticed an alignment between the semi-minor axis of NGC~5139, NGC~104, and NGC~6341 and their rotation axis, suggesting rotation as the driving mechanism of their shape. Rotation has also been detected in the GC NGC~1904, likely behind the slight flattening observed in the direction perpendicular to the rotation axis \citep{2022Leanza}.

CC10 studied the shape and orientation of 116 Galactic GCs. They found that GCs close to the Galactic bulge exhibiting a high flattening tend to have their semi-major axis pointing toward the Galactic bulge.
They concluded that tidal effects from the external Galactic potential might impact the shape of GCs located in the Galactic bulge.
CC10 compared their results with WS87, and found significant discrepancies between their results. They investigated the outliers in further detail to understand the possible cause of these differences and proposed two main interpretations. The first one is linked to the data use. CC10 used infrared observation from 2MASS covering a broader spatial extent than the optical data used by WS87. They highlighted that dust extinction could significantly alter the recovered shape of the ellipticity when using optical observations, more sensitive than infrared. The spatial extent of the available data can also play a significant role as the dense center of the cluster is expected to be rounder due to more stellar interactions than the outer regions of the cluster \citep{GD_binney_intro}. The second source of discrepancy can be linked to the different methods used. WS87 used surface photometry to compute the shape of GCs, which can be severely impacted by bright stars, while CC10 used a probabilistic star-counting technique, which could be affected by the spacing of the grid. The dependency of ellipticity measurements on the method used had already been noted by \citet{1985Fall}, corroborating the point from CC10 that bright stars might dominate the surface brightness maps, therefore leading to less reliable results than star count.

\cite{2014Fabricius} and \cite{2024Cruz_reyes} both used an eigenvector analysis of the spatial stellar distribution, with the ellipticity and position angle directly related to the eigenvalues and eigenvectors of the covariance matrix of the stellar positions. Such a method is straightforward and mathematically simple. However, it faces issues when dealing with outliers or for datasets with a low number of stars available. This can severely impact the results \citep{jolliffe2002pca}.

This overview of the investigation of flattening of GCs highlights the importance of providing a robust method capable of delivering consistent results across diverse datasets. Current approaches often depend strongly on the adopted technique, the data quality, and the number of stars available, which can lead to inconsistent ellipticity estimates. In this context, a method that remains robust even for a limited sample of stars is crucial. Such an approach will be particularly relevant for future studies focusing on the shape of multiple stellar populations (MSPs) in GCs, for instance. The origin of these MSPs remains a puzzle. Understanding their flattening patterns could provide valuable information to confirm or rule out scenarios on their formation. However, these analyses are typically restricted to a small number of stars, mostly along the red giant branch (RGB). They will require a method to measure the ellipticity that is both reliable and minimally sensitive to sample size. Providing a consistent and reproducible framework for measuring GC shapes will therefore not only clarify the role of rotation and other factors in cluster morphology but also open the way to investigating possible differences in the spatial structure of MSPs.

In this article, we present a robust method that we use to compute the shape of 29 Galactic GCs. 
In Sect.~\ref{sec:data_prep}, we describe the photometric data available. In Sect.~\ref{sec:methods}, after testing two commonly used methods, we present our robust method to recover the ellipticity of GCs, performing well on small datasets and robust to the presence of outliers. We present our results and explore the potential causes for the flattening of GCs in Sect.~\ref{sec:results}. Finally, we present our conclusions in Sect.~ \ref{sec:ccl}.

\section{Photometric data} \label{sec:data_prep}

In this study, we focus our analysis on RGB stars when computing ellipticities. This choice is motivated by the need for a stellar population with high and uniform photometric completeness across the entire field of view, including in the cluster cores, where crowding can affect the completeness.

We acknowledge that this selection significantly reduces the size of our stellar sample and excludes fainter stars that could enhance the statistical robustness of our measurements. However, the robust method presented in Sect.~\ref{subsec:robust_PCA} is corrected to perform well on small datasets (see Appendix~\ref{appendixA}). 
Dedicated tests conducted on a representative subset of clusters (see Appendix~\ref{sec:appendix_b}) indicate that ellipticity estimates using all stars might suffer from photometric incompleteness. In contrast, focusing our analysis on RGB stars yields stable, unbiased measurements.

The Stetson ground-based homogeneous catalogue \citep{2019Stetson} is one of the most extensive photometric studies publicly available. It provides homogeneous photometry in the U, B, V, R, and I filters for 48 GCs in the Milky Way. The Hubble Space Telescope UV Globular Cluster Survey catalogue (`HUGS') \citep{2015Piotto,Nardiello2018} targets the inner regions of each cluster, providing photometry in the F275W, F336W, F438W, F606W and F814W filters for 56 GCs in the Milky Way. We combined both catalogues in order to produce a wide-field view of each cluster in our sample, using the positions of all RGB stars in both catalogues to compute the ellipticity profiles of the GCs. The method for combining and cleaning the photometric catalogues is described in this section.

In order to merge the HST photometry with the ground-based catalogue in such a way that allows for detailed analysis of the overall shape of each cluster, the spatial completeness of the individual photometric catalogues was calculated. In this method, outlined in depth in Section 2.6 of \cite{2023Leitinger}, the square HST field of view was reshaped in order to smoothly transition into the ground-based photometry. To do this, we first took the original, full catalogue of HST photometry and defined annuli, spaced by $1$'', from the centre of the cluster to the outer regions. Within each annulus, $360$ artificial points were equally distributed. If an artificial point was within proximity to a real star in the photometry, that artificial point was considered to be within a spatially complete region. The first annulus to reach $50$\% spatial completeness defined the outer edge of the HST field, effectively changing the square field of view to more of a circular shape. When the ground-based photometry was then added to the HST photometry, we removed any stars in the ground-based photometry which overlapped with the HST field, ensuring a smooth transition between the HST and ground-based fields of view.

Cleaning the two catalogues included removing non-member stars that did not belong to the cluster. More specifically, non-member stars were removed with the use of proper motions provided by Gaia DR3 \cite{2016Gaia,2020Gaia}, following a method closely aligning with the one outlined in Section 2.4 of \cite{2023Leitinger}. The HST catalogues include a `membership probability' column which we used to clean the data of non-members, using a threshold of $> 75\%$. The majority of stars in the final sample of each cluster fall in the $95 - 100\%$ range with $<30$ stars per cluster showing $<95\%$ membership probability. We cross-matched the combined HST and Stetson catalogue with Gaia DR3 using TOPCAT\footnote{\url{https://www.star.bris.ac.uk/~mbt/topcat/}}. Differential reddening correction was then completed for both the HST and ground-based photometry, following the method described in Section 2.2 of \cite{2023Leitinger}, using reddening maps from \cite{2024Pancino}. To ensure the correctness of the match, we derived approximate V and I magnitudes from Gaia G magnitudes to investigate whether a roughly one-to-one correlation existed for the matched stars. Then, we used a $\chi^2$ test with the right ascension (RA) and declination (DEC) proper motions components taken from \citet{2021Vasiliev} and selected stars with $\chi^2 < 5$, which provided a reasonable balance between rejecting non-members and retaining likely cluster members. Stars with unreliable photometry in both catalogues were removed based on the sharp quality indicator, while the ground-based photometry from \cite{2019Stetson} was additionally cleaned using the $\chi$ quality indicator, as outlined in Section 2.3 of \cite{2023Leitinger}. Red giant branch stars were isolated through a combination of polynomial fitting and N-$\sigma$ clipping of the CMD in various photometric combinations outlined in Section 2.5 of \cite{2023Leitinger}. During this process, we removed horizontal branch (HB) and asymptotic giant branch (AGB) stars, as well as stars which were located too far from the RGB (i.e. blue stragglers or excessively red outliers). In terms of the photometric completeness of the two individual photometric samples, we assume 100\% completeness for HST stars at magnitudes brighter than the sub-giant branch as specified by \cite{2008Anderson}, while the Stetson catalogue is complete between magnitudes of $12 < V < 19$ mag across all radii, according to \cite{2019Stetson}. The final sample was then comprised of member RGB stars with $>50\%$ spatial completeness and reliable photometry. 

The final catalogue spans a broad spatial range: extending from 2 half-light radii in the least-covered clusters, to 23.6 half-light radii in the most well-covered ones, with a median coverage of 7.1 half-light radii. This coverage was estimated using the distance of the most distant star from the cluster center in each case, and the half-light radius from \citet{1996Harris}. The original catalogue from \citet{2023Leitinger,2024Leitinger} provides data for 30 GCs. Here, we have excluded NGC~6752 due to significant gaps in the ground-based observations from \cite{2019Stetson}, which would greatly affect the ellipticity results.

\section{Methods}
\label{sec:methods}
As mentioned in Sect.~\ref{sec:intro}, different methods to measure the shape of GCs can yield different results. 
In this section, we investigate two methods to compute the ellipticity of GCs. In the first method, described in Sect.~\ref{subsec:KDE}, we smooth the spatial distribution of stars in the GCs using a two-dimensional kernel density, and fit ellipses to different contour levels; this approach makes it possible to extract an ellipticity profile as a function of the distance to the cluster center, and is similar to the one employed by \citet{2019Stetson}. 
In the second method, presented in Sect.~\ref{subsec:PCA}, we investigate the data-driven principal component analysis (PCA) method that identifies the major and minor axes of the clusters based on the variance of the data. Each method has its advantages, assumptions, and limitations, which we further discuss in Sect.~\ref{subsec:robustness}, before presenting our robust method in Sect.~\ref{subsec:robust_PCA}.

For all the methods, our inputs are stellar positions with respect to the center of the cluster. To avoid introducing any projection effect, we converted the on-sky coordinates in Cartesian coordinates, using the formula specified in equation 1 from \citet{2006vdv}.

\subsection{Method 1: Fitting of density contours}
\label{subsec:KDE}

First, we used Kernel Density Estimation (KDE, \citealt{Wiley_KDE}) to estimate the density of stars in the Cartesian space. This non-parametric method creates a smooth and continuous function representing the density of stars.
Then, we defined a set of contour levels ranging from 10\% to 90\% of the maximum density value, and we fit the coefficients A, B, C, D, G, and H representing an ellipse described by the general conic form:

\begin{equation}
    F(x, y) = Ax^2 + Bxy + Cy^2 + Dx + Gy + H = 0 \ ,
    \label{eq:ellipse_eq1}
\end{equation}
to the stellar positions, using a Least-Squares method \citep{Halir1998NumericallySD}. Then, we can derive the ellipse's semi-major, semi-minor axes, and position angle from the coefficients.

\subsection{Method 2: Principal component analysis (PCA)}
\label{subsec:PCA}
PCA is a method aiming at reducing the number of dimensions in large datasets in order to extract the most important information \citep{Pearson1901}.
In our case, the PCA is useful to find the axes with the most information encoded.

The dataset of stellar positions is represented by the matrix $\mathbf{D}$:
\begin{equation}
\mathbf{D} = \begin{pmatrix}
x_1 & y_1 \\
x_2 & y_2 \\
\vdots & \vdots \\
x_n & y_n
\end{pmatrix} \ .
\label{eq:4_2_D}
\end{equation}
Each row \((x_i, y_i)\) represents the position of a star in a 2D Cartesian coordinate system with the origin in the center of the cluster. 
We implemented PCA using the singular value decomposition (SVD), which decomposes D into three matrices:
\begin{equation}
\mathbf{D} = \mathbf{U} \mathbf{\Sigma} \mathbf{V}^T \ .
\label{eq:4_2_SVD}
\end{equation}
In this decomposition $\mathbf{U}$ is an $N \times N$ orthogonal matrix, where $N$ is the length of the data matrix $\mathbf{D}$, in our case the number of stars in our dataset. $\mathbf{\Sigma}$ is a diagonal matrix with the singular values $\sigma_1$ and $\sigma_2$, which are related to the square roots of the eigenvalues of the covariance matrix of the data. $\mathbf{V}^T$ is a $2 \times 2$ orthogonal matrix, where each column represents the principal direction (eigenvector) of the data. The singular values $\sigma_1$ and $\sigma_2$ represent the spread of the data along each principal component. The square of each singular value gives the variance along each principal component, which is proportional to the eigenvalues of the covariance matrix. This decomposition thus provides the principal directions of the data and their associated variances.

In PCA, the eigenvector associated with the largest eigenvalue (or the largest singular value \(\sigma_1\)) is the direction along which the variance is maximized. This direction is identified as the major axis of the data distribution. Conversely, the eigenvector associated with the smallest eigenvalue (or the smallest singular value \(\sigma_2\)) corresponds to the minor axis. 
The ellipticity and the position angle are then computed as:
\begin{align}
e &= 1 - \frac{\sigma_{\text{2}}}{\sigma_{\text{1}}} \label{eq:4_2_e} \\
\Phi &= \arctan\left(\frac{v_{1y}}{v_{1x}}\right) \ ,
\label{eq:4_2_theta}
\end{align}
where \(v_{1x}\) and \(v_{1y}\) are the components of the eigenvector associated with the largest singular value (or eigenvalue), indicating the direction of maximum variance.

\subsection{Robustness and accuracy of the methods}
\label{subsec:robustness}
In this section, we will test the two different methods described above on mock GCs, varying three parameters: the number of stars $N$, the true ellipticity $e_{\rm true}$, and the true angle of the ellipse $\Phi_{\rm true}$, defined as the angle between the $x$-axis and the semi-major axis of the ellipse. We will also test the robustness of the methods with respect to the presence of outlier stars. 
These tests are particularly important to ensure that the selected method provides accurate and meaningful information on the global ellipticity of GCs.

We generated mock positions of stars following a King profile \citep{1966King}.
Then, we applied a transformation to the dataset to impose an ellipticity and an angle. Each test was realised 1000 times, on slightly different mock GCs with the same imposed ellipticity and position angle. We fixed the seed of the random generator of stars to make our tests reproducible and comparable. The mean value and the error bars on the parameters were obtained by averaging and taking the standard deviation of the realisations.

The first test that we run aims at quantifying the impact of the number of stars on the error on the ellipticity, for different values of the true ellipticity.
To do so, we uniformly sampled 300 times the number of stars $N$ between 50 and 1500, the true ellipticity $e_{\rm true}$ between 0.01 and 0.3, the true angle $\Phi_{\rm true}$ between 0$^\circ$ and 180$^\circ$, and the core radius $r_c$ between 0.5 arcmin and 15~arcmin. We used these parameters as inputs to mimic the distribution of stars in the GCs from our sample. 
We present the results in the upper panel of Fig.~\ref{fig:NKDE_NPCA}, where we plot the recovered ellipticity $e_{rec}$ as a function of the true ellipticity $e_{true}$. The points are colored according to the number of stars. In the left panel, the ellipticity was recovered using the KDE method, and in the right panel using the PCA method. In both cases, two biases are present. First, the number of stars strongly impact the error on the ellipticity. When $N<300$ (dark blue points), both methods overestimate the true ellipticity, visible as the large number of points located above the black identity line in Fig.~\ref{fig:NKDE_NPCA}.
The other source of bias comes from small ellipticities. If the cluster is close to spherical, both methods tend to overestimate the ellipticity. This is visible as larger errors for smaller $e_{true}$ in Fig.~\ref{fig:NKDE_NPCA}.
The position angle of the ellipse, not shown here, is always accurately recovered, except for very small values of the ellipticity, where the position angle is not defined.

\begin{figure*}
 \resizebox{\hsize}{!}{\includegraphics{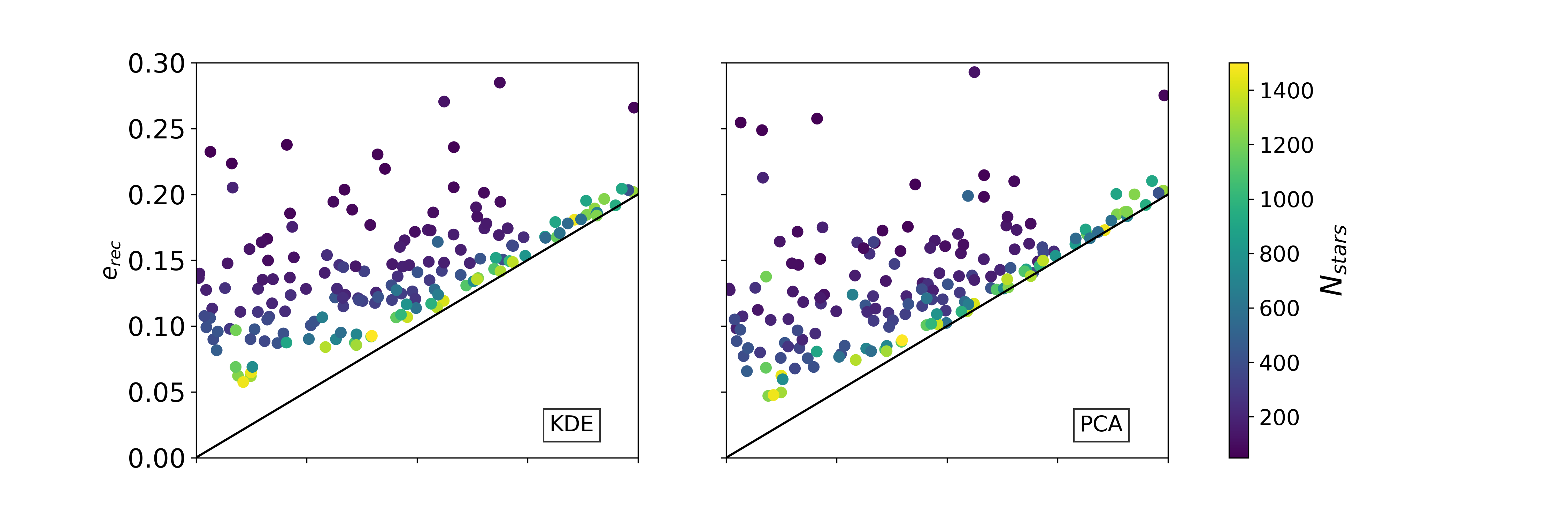}}
\resizebox{\hsize}{!}{\includegraphics{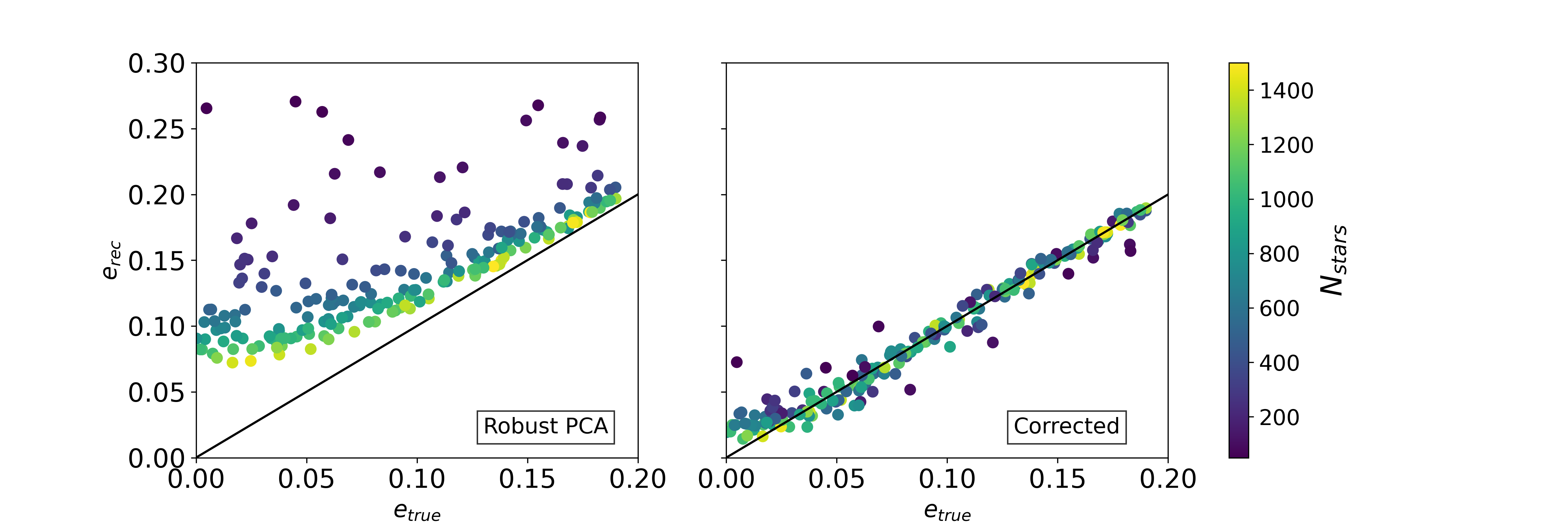}}
\caption{Recovered ellipticity as a function of the true ellipticity using the KDE method (upper-left panel), the PCA method (upper-right panel), the robust PCA method (lower-left panel) and the robust PCA method afer bias correction (lower-right panel). The points are colored according to the number of stars in the mock GCs.}
\label{fig:NKDE_NPCA}
\end{figure*}

The second test that we run is a robustness check. We manually add 10 outlier stars to mock GCs containing 1000 stars (1\% of outliers), and inspect the mean value of the ellipticity and position angle recovered using the KDE and PCA methods.
We present the results in Fig.~\ref{fig:4_kde_outliers_right} for mock GCs with $e_{\rm true} = 0.15$ and $\Phi_{\rm true} = 45^{\circ}$ (left panel), and with $e_{\rm true} = 0.25$ and $\Phi_{\rm true} = 60^{\circ}$ (right panel). 
The values for $N$ and $e_{true}$ have been chosen to avoid including any of the biases presented above.
We run this test with two types of outliers, the first ones are added in the direction of the semi-major axis (yellow points), and the second ones are added in the direction of the semi-minor axis (pink points).
In the first case, adding 1\% of outlier stars has a strong impact on the recovered ellipticity regardless of the method used and of the value of the true parameters. The ellipticity is overestimated due to the presence of these outliers along the semi-major axis.
In the second case, when outliers are added along the semi-minor axis, the recovered position angle is shifted by 10° to 90°, and the ellipticity is underestimated by 0.05 to 0.2, depending on the method. This means that outliers distributed perpendicular to the intrinsic flattening tend to rotate the recovered major axis and make the cluster appear rounder than it actually is.

These tests on mock data suggest the necessity to develop a robust and accurate method to measure the ellipticity in GCs. This will be very important when measuring the ellipticity of clusters where the number of stars available is low, for example, when studying the shape of multiple stellar populations in GCs.
One of the main advantages of the PCA method is its speed. Therefore, we decided to adapt the PCA to make it robust to outliers and to correct for the biases present in small datasets and for round clusters. 

\begin{figure}
\resizebox{\hsize}{!}{\includegraphics{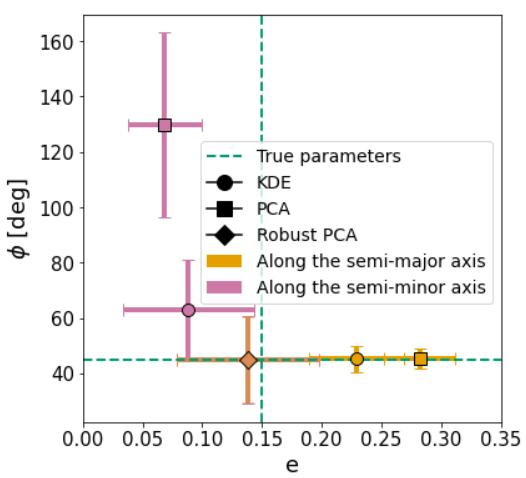}}
\resizebox{\hsize}{!}{\includegraphics{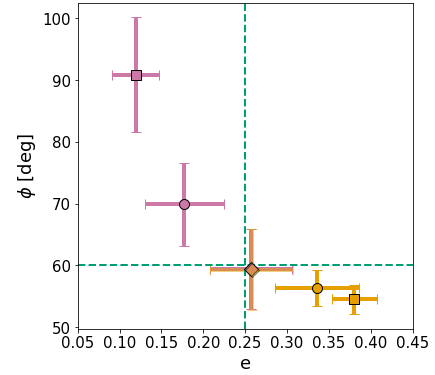}}
\caption{Impact of outliers on the recovered ellipticity and position angle for the KDE (circles), PCA (squares), and robust PCA (diamonds) methods for two sets of true parameters. The colour of the points indicates the location of the outliers, in pink with positive x and negative y, corresponding approximately to the direction of the semi-minor axis and in orange with positive x and positive y, corresponding approximately to the direction of the semi-major axis of the ellipse. Outliers were added within fixed quadrants relative to the cluster centre, not strictly along the ellipse axes, which leads to small asymmetries in recovered angles when the ellipse position angle changes. The green dashed line represents the true value of the ellipticity and angle, taken to 0.15 and 45° (left panel) and to 0.25 and 60° (right panel).  We note that, for the Robust PCA method, only one point is visible, thus highlighting the robustness of the method. Regardless of the direction where outliers are added, the ellipticity and position angle are accurately recovered.}
\label{fig:4_kde_outliers_right}
\end{figure}

\subsection{Robust PCA method and bias correction}
\label{subsec:robust_PCA}
The robust PCA uses a distinct approach to compute the covariance matrix, relying on the Minimum Covariance Determinant method (MCD, \citealt{Rousseeuw1984}). This method identifies a subset of the data with the smallest possible covariance determinant, which measures the spread of the distribution. By minimizing this determinant, the method selects the subset of data that is most tightly clustered, thus reducing the influence of outliers on the final covariance estimate. We use the scikit-learn class MinCovDet\footnote{\url{https://scikit-learn.org/stable/modules/generated/sklearn.covariance.MinCovDet.html}} to compute the covariance matrix using the MCD method. Then, as done in Sect.~\ref{subsec:PCA}, we compute the eigenvalues and eigenvectors of the covariance matrix and estimate the ellipticity and angle of the data.

To test the robustness of this method, we perform the same test as in Sect.~\ref{subsec:robustness}. We add the ellipticities recovered using the robust PCA in Fig.~\ref{fig:4_kde_outliers_right} as diamonds.
The recovered ellipticities and angles are almost superimposed on the lines indicating the true values of these parameters. This agreement holds regardless of the directions in which outliers were added. This method is, thus, very robust to the presence of outliers. 
However, we find that it is still sensitive to the number of stars, as attested by the lower left panel of Fig.~\ref{fig:NKDE_NPCA}, showing the recovered ellipticity $e_{rec}$ as a function of the true ellipticity $e_{true}$. where points are colored according to the number of stars. Therefore, another correction should be applied. We constructed a bias correction function using a set of mock data to correct for the systematic biases linked to small datasets ($N \simeq 100$) and to the overestimation of the ellipticity.
More information on the bias correction performed is available in Appendix~\ref{appendixA}. The correction of the bias function is visible in the lower-right panel of Fig.~\ref{fig:NKDE_NPCA}. After applying the correction, most of the points fall close to the 1:1 black line. The bias linked to a small number of stars is efficiently tackled.

\section{Ellipticity of 29 MW GCs} 
\label{sec:results}
To compute the global ellipticity and position angle, we randomly sample 90\% of all the stars in each GC 1000 times. We then use the mean value and standard deviation of the ellipticities and position angles obtained for these realisations as measurements and corresponding error estimates for these quantities.
\subsection{Comparison with literature}
\label{subsec:literature}
In this section, we present the ellipticity and position angle values obtained for the 29 Galactic GCs in our sample by using the robust PCA method described in Sect.~\ref{subsec:robust_PCA}. For clusters with $N_{stars}<1500$, we further applied a correction model to correct for the biases linked to small datasets (see Appendix~\ref{appendixA}).
Then, we compare these values with those from WS87, CC10, and CR24.

In our analysis, we will refer to GCs as round or spherical when they have ellipticities $e<0.05$, and as flattened or elliptical otherwise.
According to this criterion, among our 29 GCs we find 16 flattened GCs.

\begin{figure*}[t]
\centering
\includegraphics[width=\textwidth]{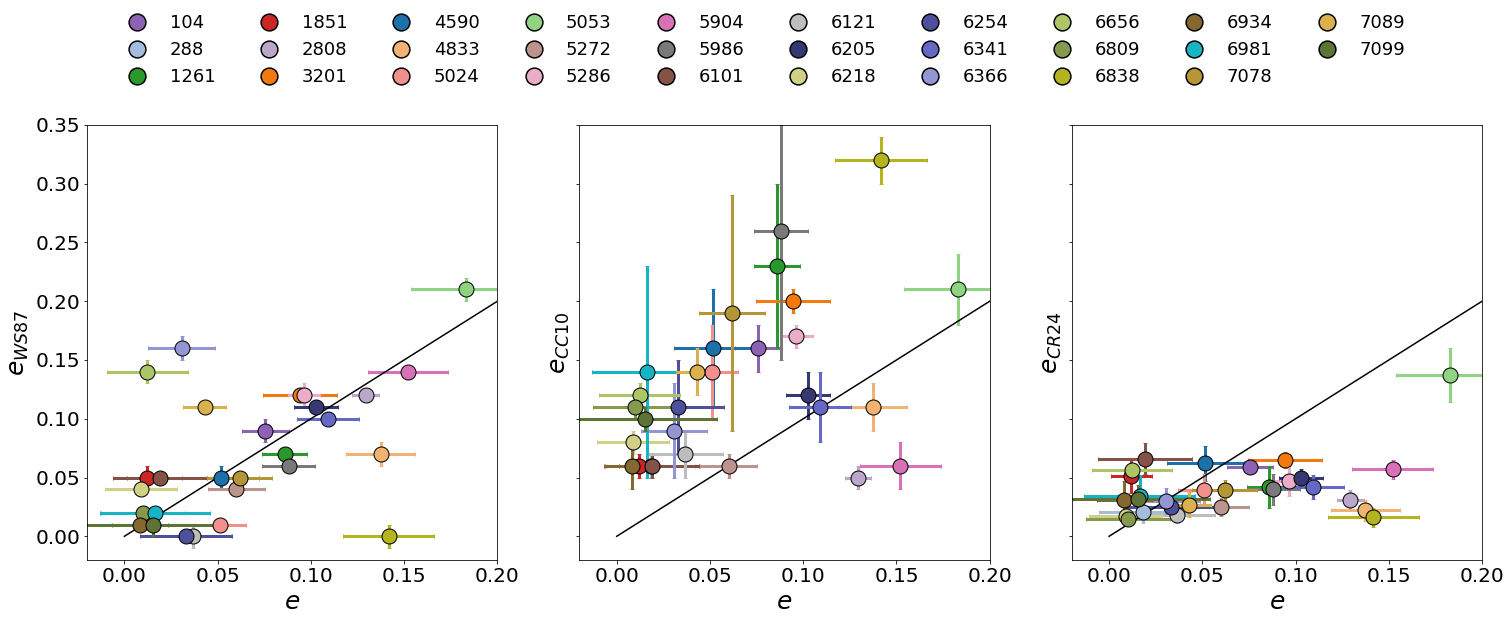}
\caption{The ellipticities of the 29 Galactic GCs in our sample, measured by using the robust PCA method, are here compared with the estimates from the literature, provided by \citet{1987White} (left panel), \citet{Chen_2010} (middle panel), and \citet{2024Cruz_reyes} (right panel). The solid lines indicate equal ellipticities; each GC is represented by a coloured point (see legend on top), and error bars are indicated.}
\label{fig:4_etot_comparison}
\end{figure*}

The results of our analysis are presented in Table~\ref{tab:cluster_results}, where we provide the values of ellipticity $e$, together with ellipticity measured by WS87, CC10, and CR24 ($e_{\rm WS87}$, $e_{\rm CC10}$ and $e_{CR24}$, respectively).
We compare our results with WS87, CC10, and CR24 in the left, middle and right panels of Fig.~\ref{fig:4_etot_comparison}, respectively.
Overall, our values show better agreement with WS87, with reasonable consistency within 3$\sigma$, except for five GCs, namely NGC~6981, NGC~6838, NGC~6366, NGC~6656, and NGC~7089. Particularly striking is the situation of NGC~6838, NGC~6656 and NGC~6366: the first appears to be round in WS87, but we find it to be very flattened, and for the second and third, the situation is reversed. 
The analysis by CC10 resulted in higher ellipticity values in general, which differ from our findings. This discrepancy may arise because their measurements are more sensitive to bright stars and the outer regions of globular clusters, where flattening tends to be more pronounced.
The case of NGC~6838 is once again particularly interesting, as CC10 found it very flattened, while WS87 found it round; our measurement appears to be somewhat in between those estimates. We will further discuss these results in Sect.~\ref{subsec:group2}.

Finally, CR24 computed the ellipticity of 163 globular clusters using the on-sky distribution of cluster members and the PCA method. We compare our results with theirs in the right panel of Fig.~\ref{fig:4_etot_comparison}.
Our results differ for the majority of clusters we have in common, as we find higher values of ellipticity for 14 out of 26 clusters.
These discrepancies can be linked to two points.
First, CR24 used the right ascension and declination of stars for the PCA analysis, while we used the Cartesian coordinates. Computing the covariance matrix from stellar positions in RA and Dec will indubitably generate projection effects and errors in the distance estimate, which will propagate to the measurement of ellipticity. 
Second, as demonstrated in Sect.~\ref{subsec:robustness}, only a robust implementation of the PCA can prevent the sensitivity to outliers. So if outliers are present in the data used by CR24, this will impact the resulting ellipticities. This issue also applies to methods based on fitting density contours, although the impact of outliers is generally less visible than in the case of PCA (see Fig.~2).

\subsection{The role of relaxation time, rotation, velocity anisotropy, and tides}
\label{subsec:role}
In this section, we explore the factors that can influence the shapes of GCs. 

The first quantity that we investigate is the relaxation time. It can be defined as the timescale on which encounters between stars, leading to energy exchange, produce major changes in the structure of the cluster, without disturbing its dynamical equilibrium \citep{1997Meylan}. 
Our results are shown in Fig.~\ref{fig:4_phi_etot}, where we put into relation the ellipticity that we measure for the 29 MW GCs in our sample and the relaxation time at the half-mass radius taken from \citet{1996Harris}\footnote{The relaxation time is taken from the updated Harris catalogue available online: \url{https://physics.mcmaster.ca/~harris/Databases.html}}. The GCs are colored according to the time of the last disk crossing from \citet{2024Pancino}.
A trend is visible in the figure: clusters with longer relaxation times (dynamically young) are, in general, more elliptical, and those with short relaxation times (dynamically old) are more round. An evident exception is NGC~6838, which is very flattened ($e = 0.14$), even if it appears to be dynamically old. As mentioned in Sect.~\ref{subsec:literature}, its high flattening might be related to its recent crossing through the Galactic disk.

\begin{figure}
\resizebox{\hsize}{!}{\includegraphics{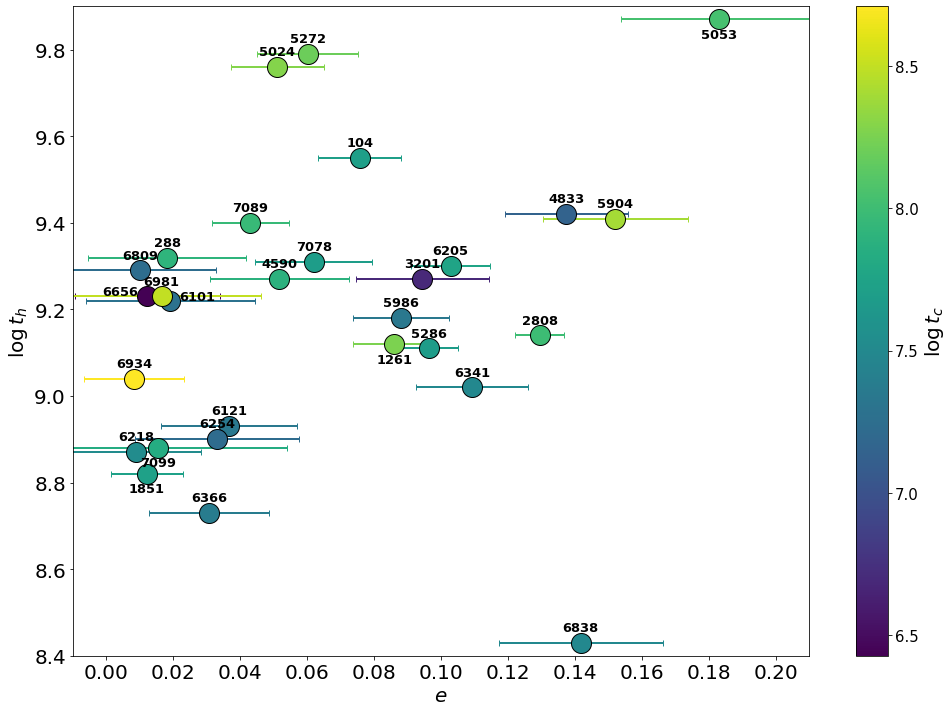}}
\caption{The ellipticity $e$ of 29 Galactic GCs, obtained using the robust PCA method, is shown here against the relaxation time at the half-mass radius from \citet{1996Harris}. The points are colored according to the last crossing time through the Galactic disk, taken from \citet{2024Pancino}. The name of each cluster is indicated close to the corresponding point. Both times are given in years. Dynamically young clusters tend to be more elliptical.}
\label{fig:4_phi_etot}
\end{figure}

These findings indicate the importance of understanding the origin of the deviation from spherical symmetry of GCs.
To identify the primary driver behind the observed flattening in GCs, whether it is internal rotation, velocity anisotropy, or tidal effects, we first consider the role of rotation.

The $V/\sigma$ diagram is a tool frequently used to study elliptical galaxies \citep{2007Cappellari,Emsellem2011,2017Brought}. The relation between projected rotation signature and flattening is derived from the tensor virial theorem \citep{1978Binney} and for an edge-on system is given by:
\begin{equation}
    \left(\frac{V}{\sigma}\right)_{ed}  =k_1 \sqrt{\frac{e}{1-k_2e}} \ ,
\label{eq:4_Vsig_e}
\end{equation}
where the subscript $ed$ indicates the edge-on case, $V$ is the maximum rotation velocity along the line-of sight, $\sigma$ is the central velocity dispersion, $e$ is the ellipticity, and $k_1$ and $k_2$ are constants that account for the intrinsic properties of the system, including its geometry and kinematics. In our analysis, we assume $k_1 = 0.831$ and $k_2 = 0.896$ \citep{2007Cappellari,2013Bianchini}

Such a tool has already been applied to GCs to understand the link between flattening and rotation or anisotropy. Using this diagram with ellipticities from WS87, \citet{2014Fabricius} concluded that rotation is the driving factor of the ellipticities, as higher flattening exhibits higher $V/\sigma$. However, their analysis did not consider projection effects, which have a significant impact when $i<30$°. 
For example, \citet{2013Bianchini} investigated three clusters: $\omega-Cen$, NGC~104 and NGC 7078. By placing them in the $V/\sigma$ diagram and correcting for their inclination, they showed that the positions of all three clusters were shifted, demonstrating the impact of projection effects. They highlighted that deviation from the line of isotropic oblate rotators could be a hint that several mechanisms are acting together, such as inclination, differential rotation, and pressure anisotropy.
\citet{2018Kamann} found a similar trend in this diagram; increasing flattening with higher rotational signature, though with a lower significance, using ellipticities from WS87.

The quantities presented in Eq.~\ref{eq:4_Vsig_e} need to be corrected to account for the inclination of the GCs. This can be done as follows \citep{1987Binney}:
\begin{align}
    \left(\frac{V}{\sigma}\right)_{ed} &= \left(\frac{V}{\sigma}\right)_{obs} \frac{\sqrt{1 - \delta \cos^2i}}{\sin i} \label{eq:4_vsig_corr} \\
    e_{\rm intr} &= 1 - \sqrt{1 + e(e-2)/\sin^2 i} \label{eq:4_e_corr} \\
    \delta &= 1 - \frac{1+ (V/\sigma)^2}{[1-\alpha(V/\sigma)^2]\Omega(e_{\rm intr})} \label{eq:delta},
\end{align}

where $i$ represents the inclination angle, $\delta$ the anisotropy and $\Omega(e_{\rm intr})$ is a function of the cluster's intrinsic ellipticity $e_{\rm intr}$ or inherent shape, excluding projection effects, provided in Eq.~16 of \citet{2007Cappellari}. $\alpha$ is a dimensionless parameter depending on the streaming velocity and the stellar density. Similarly to \citet{2007Cappellari}, we adopt a fixed value of $\alpha=0.15$. For the small ellipticities and rotation strengths relevant to our clusters, varying $\alpha$ has a negligible effect on the results (see Fig.~2 of \citealt{Binney2005}).

\begin{figure*}
\centering
\includegraphics[width=17cm]{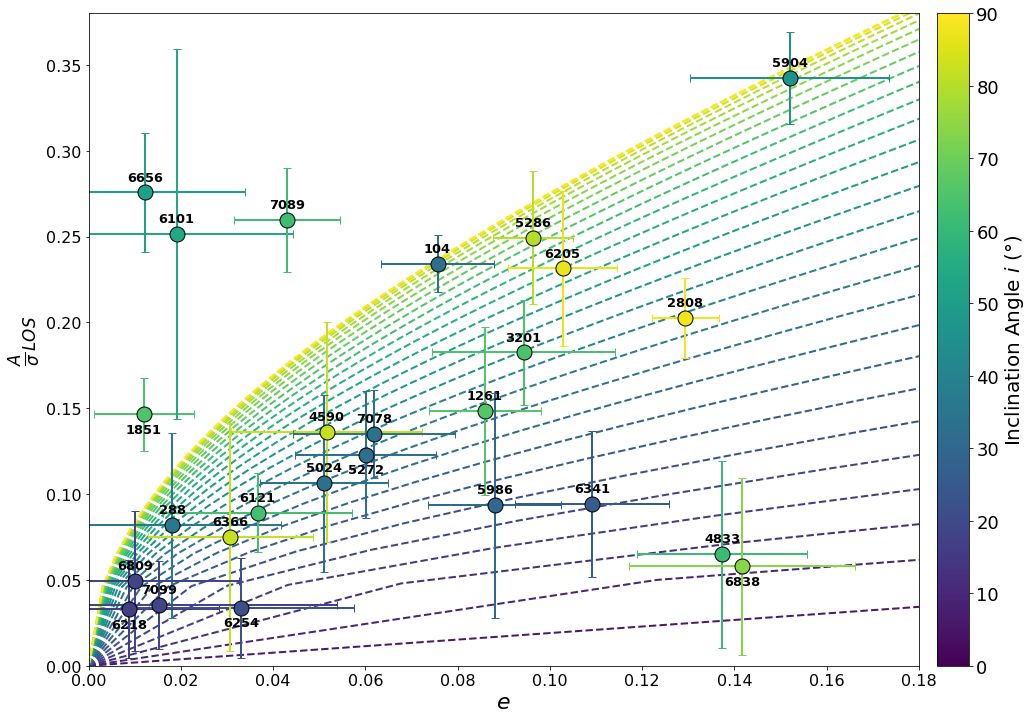}
\caption{This plot shows the rotation strength, taken from \cite{2024Leitinger}, as a function of the flattening we measured. The names of GCs are indicated close to the corresponding points, colored according to the inclination angle of the cluster, also taken from \cite{2024Leitinger}. Low inclinations are plotted in blue while high inclinations are colored in yellow. The uncertainties associated with the inclination angles are not shown in this figure. Instead, we report them in the last column of Table~\ref{tab:cluster_results}.
The dashed lines indicate the relation for isotropic oblate rotators viewed at different inclination angles.}
\label{fig:v_sigma_e}
\end{figure*}

In our analysis, we use the amplitude $A$ of the best-fit rotation curve as a proxy for the projected rotational velocity $V$, as commonly adopted in past studies \citep{2013Bianchini, 2018Kamann}. We present our results in Fig.~\ref{fig:v_sigma_e}, where we plot our ellipticity values and the order-to-random motion parameter $A_{LOS}/\sigma$ from \citet{2024Leitinger}. Three clusters, NGC~5053, NGC~6934, and NGC~6981, lack a $A_{LOS}/\sigma$ measurement and are therefore not included in the plot.
The yellow line indicates the theoretical relation for an isotropic oblate rotator viewed edge-on (Eq.~\ref{eq:4_Vsig_e}). The other colorful lines show the relation between $V/\sigma$ and $e$ when $i \neq 90$°, as described by Eqs.~\ref{eq:4_vsig_corr}, \ref{eq:4_e_corr}, and \ref{eq:delta}. Each point is colored according to the inclination angle of the GC. Inclinations in \citet{2024Leitinger} are reported over 0–180°, but for our analysis we adopt the standard 0–90° convention\footnote{For inclination angles greater than 90°, we apply the transformation $i_{corr} = 180 - i$. Correspondingly, because folding the inclination beyond 90° reverses the rotation axis direction, we adjust the position angle by adding 180° to the rotation angle when the original inclination exceeds 90°.}. \citet{2024Leitinger} inclination angles are reported in the last column of Table~\ref{tab:cluster_results}.
Therefore, each GC should lie in the corresponding coloured line if they behave as isotropic oblate rigid rotators. 
While most of the clusters follow the line corresponding to their inclination in the framework of the isotropic oblate rotator, a few clusters deviate from their expected location. To further analyse these results, we divided the clusters into three groups. In the first group, we gather the GCs with a flattening coherent with their rotational signature (Sect.~\ref{subsec:group1}). In the second group discussed in Sect.~\ref{subsec:group2}, we include the clusters more flattened than expected. Finally, the third group gathers clusters less flattened than expected (Sect.~\ref{subsec:group3}).
All clusters have uncertainties associated with their inclination angles, also reported in Table~\ref{tab:cluster_results}. These uncertainties should be taken into account in the interpretation of the results, as large uncertainties can lead to a wide range of values for $\sin i$ in Eqs.~\ref{eq:4_vsig_corr} and \ref{eq:4_e_corr}, resulting in variations in the intrinsic rotation strength and flattening, therefore challenging the interpretation of the results. The most affected clusters are those with low inclination and large uncertainties, such as NGC~6838.

\subsubsection{Group 1: GCs with a flattening coherent with their rotational signature}
\label{subsec:group1}
The majority of the GCs appear to have the ellipticity and rotation strength consistent with what is expected in the isotropic oblate rotator framework, given their inclination. This is the case of the spherical and low or non-rotating clusters: NGC~288, NGC~6218, NGC~6254, NGC~6366, NGC~6809, and NGC~7099. 

Among the GCs with a flattening coherent with their rotational signature, eleven are elliptical: NGC~1261, NGC~3201, NGC~4590, NGC~5024, NGC~5272, NGC~5286, NGC~5986, NGC~6121, NGC~6205, NGC~6341, and NGC~7078.  Their ellipticity is in good agreement with the one expected in the isotropic oblate rotator framework, given their rotational signature, indicating the major role of rotation in shaping these clusters. As further confirmation that rotation is the main driver of the flattening in these clusters, we analyse the orientation of their flattening axis and their rotation axis. 
In Fig.~\ref{fig:4_phi_rot_tot}, we show the position angles of the ellipse we measure, $\Phi$, and compare them to the rotation axis position angles, $\theta_0$, from \citet{2024Leitinger}. 
To compare our measured ellipse position angles with the rotation axis position angle, we account for the difference in reference frames. Our PCA-derived angles are measured counter-clockwise from the west direction, whereas the angles in \cite{2024Leitinger} are measured counter-clockwise from north. To bring both measurements to the same reference, we apply a 90° shift to their angles. After this adjustment, we wrapped the rotation angles to the [0°, 180°) interval. We also subtracted 90° from the ellipse's position angle to plot the semi-minor axis instead of the semi-major axis, allowing for a direct comparison with the rotation angle.
Here, we only consider the clusters flagged as elliptical ($e>0.05$). Non-rotating clusters with a measured $A/\sigma$ in \citet{2024Leitinger} are plotted with a star symbol. Their rotation angle might be unreliable. The black line indicates the identity: a cluster falling on this line would suggest that its flattening could be due to rotation, as the semi-minor axis of the cluster aligns with the axis of rotation. Five clusters show position angles consistent with the identity line within $1\sigma$, suggesting a strong alignment between the rotation and flattening axes: NGC~3201, NGC~5986, NGC~6205, NGC~6341, and NGC~7078. That is also the case of NGC~2808 and NGC~104, but their position in the $V/\sigma$ diagram prevents us from a definitive conclusion. Their case will be examined in more detail in Sect.~\ref{subsec:group2} and \ref{subsec:group3}. We found that the geometric position angles and rotational position angles are consistent within 2$\sigma$ for the clusters NGC~5024, NGC~5286, and NGC~5904.
We conclude that the flattening in these eight clusters is mainly due to rotation.
Finally, the position angle of the rotation axis and the semi-minor axis of the ellipse differ from the identity for NGC~5272. However, this cluster still lies along the isotropic oblate rotator relation in the $A/\sigma$–$e$ diagram, suggesting that its flattening is primarily driven by rotation, and that the observed misalignments may result from projection effects rather than anisotropy.

\begin{figure}[t]
\centering
\includegraphics[width=0.49\textwidth]{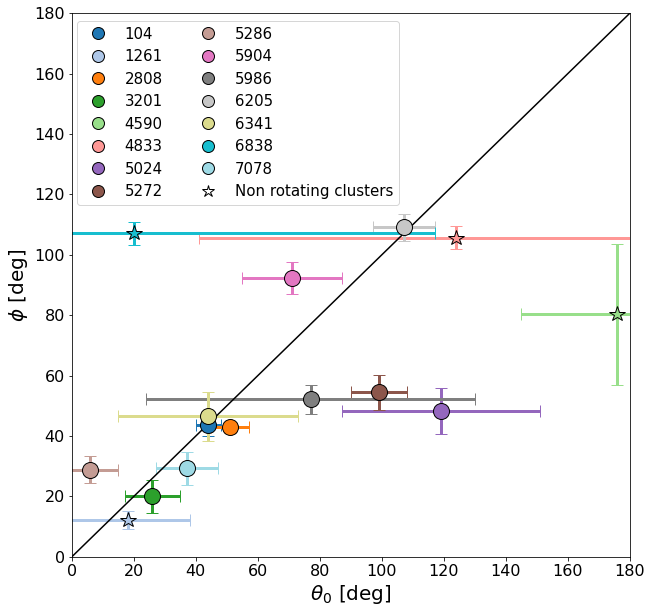}
\caption{Comparison between the position angle of the rotation axis from \citet{2024Leitinger} and the position angle measured in this work. The black line indicates the identity: in this case, this represents an alignment between the semi-minor axis of the ellipse and the rotation axis of the cluster. We only include here the clusters classified as flattened and with available position angle for the rotation axis. Some clusters are classified as non-rotating in \citet{2024Leitinger}, but they still have a measured rotation strength and position angle for the rotation axis. They are plotted with a star symbol.}
\label{fig:4_phi_rot_tot}
\end{figure}

\subsubsection{Group 2: Clusters which are more flattened than expected}
\label{subsec:group2}
Three clusters are located well below their expected position in Fig.~\ref{fig:v_sigma_e}, with a much lower rotation strength ($A/\sigma$) than expected for their ellipticity: NGC~2808, NGC~4833, and NGC~6838.

Many studies have confirmed rotation in NGC~2808 \citep{2012Bellazzini,2015Lardo,2017Jeffreson,2018Kamann,2019Sollima,2023Martens}. Gaia DR2 observations have revealed a population of extra-tidal candidate stars. These stars show a clear over-density in the trailing direction of the cluster's orbit, a feature commonly associated with tidal stripping and disc-shocking events \citep{2021Kundu}. Deep DECam imaging further revealed a substantial population of stars beyond the King tidal radius, as well as the presence of extended tidal tails with complex morphology \citep{Carballo2017}. These findings suggest that NGC 2808 resides in a dynamically rich region of the Galaxy and has likely undergone significant tidal interactions, which may have contributed to its enhanced flattening. As we find an excellent agreement between the ellipse minor axis and the rotation axis of the cluster in Fig.~\ref{fig:4_phi_rot_tot}, we conclude that the high flattening of NGC~2808 is likely a combination of internal rotation and tidal interactions.

NGC~6838 is a non-rotating cluster \citep{2018Bianchini,2019Vasiliev,2024Leitinger}. It is dynamically old, therefore isotropy throughout the cluster is expected as a natural consequence of dynamical relaxation. 
We find it to have a significant flattening. This suggests that additional mechanisms like tidal interactions play a role in determining its shape.
\citet{Chen_2010} suggested that the high flattening of NGC~6838 might be related to its recent crossing through the Galactic disk some 16~Myr ago \citep{2008VandePutte}, much shorter than the relaxation time of the cluster, which is 270~Myr \citep{1996Harris}. More recent estimates of the time since the last galactic disk crossing have been computed by \citet{2024Pancino}, using proper motion based on Gaia DR3. They found a last disk crossing time of 31.2 Myr for NGC~6838, almost twice the value from \citet{2008VandePutte}. 
If the cluster was affected by disk shocking during this passage, we might also expect some stripped stars to remain in its vicinity. Although \citet{2021Ibata} did not list NGC~6838 among clusters with detected extra-tidal structures, this is likely due to its position at low Galactic latitude, where stream detectability is more challenging. Disk shocking could also disperse stars more diffusely, preventing coherent structures like stellar streams from forming. \citet{Chen_2010} identify some extra-tidal member stars with a more clumpy spatial distribution, and \citet{Ferrone_2023} predict that NGC~6838 should experience substantial mass loss due to repeated disk crossings. Given its present low mass ($\sim 5\times 10^4$~M$_\odot$, \citealt{2018Baumgardt}), it is plausible that this process has already removed a significant fraction of its original stellar content.
We therefore consider the remnant tidal effects from disk crossings to be a compelling explanation for the apparent flattening of NGC~6838, supported by its orbital history, low mass, and indirect evidence for past mass loss.

The case of NGC~4833 is particularly interesting. We find a flattening near 0.14, in good agreement with CC10, but the cluster is non-rotating \citep{2015Lardo,2018Bianchini,2019Vasiliev,2024Leitinger}. So rotation cannot account for its ellipticity. This cluster has a last crossing time through the Galactic disk among the shortest in our sample, as shown by its dark blue colour in Fig.~\ref{fig:4_phi_etot}. A reason for its high flattening could be tidal effects from this last disk crossing. However, the results on the ellipticity using all stars are greatly different for this specific cluster, as reported in Appendix~\ref{sec:appendix_b}. Dedicated simulations would be valuable to further investigate this complex case.

While most clusters have uncertainties on their inclination angle which do not significantly impact the results and their interpretation, NGC~4833 and NGC~6838 both exhibit large uncertainties: the inclination angle of NGC~4833 could vary between 35° and 87° (almost edge-on), and between 4° (face-on) and 64° for NGC~6838. This wide range results in a large variation of $\sin i$, leading to significant potential variations in the intrinsic flattening and rotational strength. Therefore, the interpretation for these clusters should be considered with caution due to these large uncertainties.

\subsubsection{Group 3: Clusters which are less flattened than expected}
\label{subsec:group3}
NGC~104, NGC~1851, NGC~5904, NGC~6101, NGC~6656, and NGC~7089 are located above the isotropic rotator line corresponding to their inclination, as shown in Fig.~\ref{fig:v_sigma_e}, suggesting that they are less flattened than expected given their rotation. This might suggest that another (or more) mechanisms are contributing to the cluster shape and counteracting the effect of rotation.

NGC~104 and NGC~5904 both lie close to the edge-on line, above their expected location. Inspecting their position in Fig.~\ref{fig:4_phi_rot_tot}, the semi-minor axis of NGC~104 is perfectly aligned with its rotation angle. It is also the case of NGC~5904 within 1.5$\sigma$. This suggests rotation as the primary driver of the flattening in these two clusters, with a secondary mechanism partially counteracting this effect. NGC~104 exhibits radial anisotropy in the outer regions of the cluster, defined as the intrinsic random motion between stars, as opposed to the anisotropy induced by the fast rotation of the cluster \citep{2018_Milone_47tuc,2020Cordoni,2022Libralato,2024Leitinger}. This leads us to conclude that the flattening of NGC~104 could be linked to both rotation and radial anisotropy. We note that our results for this cluster are in very good agreement with \citet{2013Bianchini}. 
NGC~5904 exhibits either isotropy \citep{2022Libralato} or mild tangential anisotropy \citep{2024Leitinger}. Its most recent passage through the Galactic disk is among the oldest of the globular clusters in our sample. Neither anisotropy nor tidal effects seems to account for the shape of the cluster. Given that its position in the $V/\sigma$ diagram is consistent within two sigma with the isotropic oblate rotator line at the corresponding inclination angle, and that the semi-minor axis aligns closely with the rotation axis, we conclude that rotation is the primary driver of the cluster’s flattening.

NGC~1851 is a dynamically old cluster. Several studies have detected rotation within the cluster \citep{2012Bellazzini, 2015Lardo, 2018Kamann, 2023Martens, 2024Petralia, 2024Leitinger}, while others have not found evidence for it \citep{2018Bianchini, 2019Vasiliev, 2019Sollima}. The cluster is characterised as isotropic \citep{2024Leitinger}. Given that the cluster's position in Fig.~\ref{fig:v_sigma_e} agrees within $2-\sigma$ with the expected location in the context of the isotropic oblate rotator, and that discrepancies have been found regarding the rotation pattern, we conclude that there is no conflict with our results.

\citet{2024Leitinger} detected for the first time a low rotation signal in NGC~6101. They also highlighted the presence of radial anisotropy in the outer regions of the cluster. The cluster exhibits peculiar dynamics, characterised by the absence of mass segregation \citep{2015Dalessandro} and a moderate likelihood of hosting a large population of black holes \citep{2016Peuten,2018Askar,2020Weatherford}. Tidal tails have also been detected \citep{2021Ibata}. The large error bars associated with the rotation strength and the ellipticity, combined with a complex internal dynamics and tidal interaction events, lead us to conclude that several parameters are likely to contribute to the shape of NGC~6101.

NGC~6656 is a well studied GC \citep{2010Lane,2018Kamann,2018Bianchini,2019Vasiliev,2019Sollima,2024Leitinger}. It exhibits a high rotational signature and a relatively small ellipticity.
We note that both WS87 and CC10 found an ellipticity greater than 0.1 for this cluster, which would shift its position to the right in Fig.~\ref{fig:v_sigma_e}, making it compatible with the results expected in the framework of the isotropic oblate rotator. The last galactic disk crossing time has been estimated to be 2.7~Myr  \citep{2024Pancino}, placing NGC~6656 to be the latest GC in our sample crossing the Galactic disk. This cluster is a good candidate in the search for tidal tails \citep{2014Kunder}. It also hosts groups of stars with differences in heavy elements abundances \citep{2009Marino,2015Mucciarelli}, suggesting a potential extragalactic origin or that it is a remnant of a disrupted dwarf galaxy \citep{2009Dacosta}. However, according to its kinematics, NGC~6656 is expected to be an in-situ cluster born in the MW main disk \citep{2019Massari}. In this context, the complex dynamics and history of NGC~6656 could explain its deviation from the isotropic oblate rotator framework exhibited in Fig.~\ref{fig:v_sigma_e}.

Significant rotation has also been detected in NGC~7089 \citep{2015Kimmig,2018Kamann,2018Bianchini,2019Vasiliev,2019Sollima,2023Martens}, as well as tidal tails \citep{2022Grillmair}. The cluster exhibits mild radial anisotropy in the outer regions \citep{2024Leitinger}. A possible explanation for its low flattening and high rotation could be the preferential stripping of stars on radial orbits during tidal interactions, as attested by the presence of tidal tails. This could lead to a rounder appearance of the cluster, despite the significant internal rotation.

\subsection{Impact of radial coverage}
\label{subsec:radial_coverage}
In the results presented in Sect.\ref{subsec:role}, the ellipticity has been computed using all the RGB stars available in our sample. This results in an inhomogeneous radial coverage between clusters, with some clusters being covered up to ~2 $ r_h$ (e.g., NGC~6121, NGC~6656) and others up to 19 $r_h$ (e.g., NGC~1261, NGC~2808). To evaluate the impact of this inhomogeneous radial coverage, we restrict the data to 3$r_h$, and remove all the stars with a distance to the cluster centre superior to 3$r_h$. We note that such a cut can also introduce its own bias, toward lower ellipticity as the centres of clusters lose their initial conditions faster than their outer regions. We plot in Fig.~\ref{fig:e_versus_e3RH} the total ellipticity $e_{tot}$, as reported in the results from Sect.~\ref{subsec:role}, versus the ellipticity within 3$r_h$. We remove NGC~6121 and NGC~6656 from the sample, as their radial coverage extends only to 2$r_h$. We highlight the names of the clusters for which the ellipticity results deviate by more than $1.5 \sigma$, indicating a significant discrepancy beyond expected uncertainties. NGC~104, NGC~1261, NGC~6341 all have an ellipticity within $3r_h$ significantly lower than when using all the stars. The two latest are among the clusters with the most extensive radial coverage. This difference could be attributed to the fact that the inner regions of clusters tend to relax faster than their outer parts. Consequently, restricting the analysis to within $3r_h$ excludes the outer, more elliptical regions, leading to a lower overall measured ellipticity.
Regarding NGC~104, the analysis from Sect.~\ref{subsec:group3} revealed that a mechanism is likely counter-acting the effect of rotation. The ellipticity within 3$r_h$ leads to a similar conclusion.

NGC~5986 is the only cluster with a statistically significant higher ellipticity within 3$r_h$. This is peculiar, as we expect the centre of a cluster to relax faster and the outer parts to be more subject to tidal effects, potentially making them more elliptical. \cite{2018Lanzoni_5986} detected peculiar kinematics for this cluster, with a solid-body rotation pattern, and a velocity dispersion profile flattening after one half-mass radius. In this scenario, the inner regions of the cluster could preserve the effects of rotation more strongly. When stars at large radii are included in the analysis, their isotropic kinematics could dilute the rotation-induced flattening in the central parts, resulting in a lower ellipticity. Such a result remains compatible with the analysis from Sect.~\ref{subsec:group1}, which shows that NGC~5986's high flattening is likely due to rotation.

We therefore conclude that the interpretation of the ellipticity results discussed in Sect.~\ref{subsec:role} remains robust when adopting a homogeneous radial coverage between clusters.

\begin{figure}[t]
\centering
\includegraphics[width=0.49\textwidth]{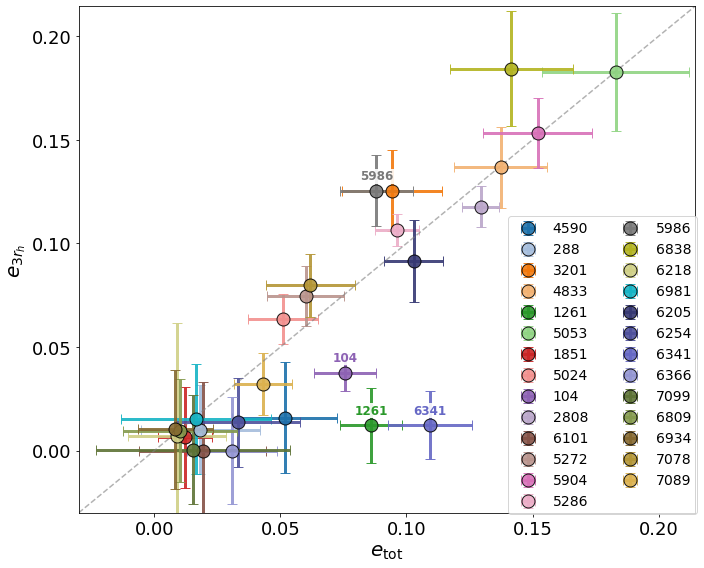}
\caption{Ellipticity computed using the robust PCA method with all stars (x-axis) and with all stars within 3$r_h$ (y-axis). Only four clusters show ellipticities incompatible within their respective error bars.}
\label{fig:e_versus_e3RH}
\end{figure}
 
\section{Conclusion} \label{sec:ccl}  
We have used photometric ground-based observations \citep{2019Stetson} combined with HST data \citep{Nardiello2018,2015Piotto} to model the flattening of 29 Galactic GCs.

To do this, we have implemented and investigated two methods that provide a measurement of the flattening and of the position angle of the elliptical shape representing the cluster. The first method relies on a kernel density estimate of the distribution of stars and an ellipse fitting to different isodensity contours. This approach has the advantage of providing an ellipticity profile, and is particularly interesting for clusters for which we expect a variation of the flattening with the radial distance. However, the method is sensitive to outliers, which can considerably alter the results. 
The second method investigated is PCA, which captures the direction along which the variance of the data is minimum and maximum, providing very fast results, but is sensitive to outliers.
The two methods that we have tested are not robust and not accurate, as they systematically overestimate small ellipticities, leading to spurious flattening measurements. 
Therefore, we further developed and improved the fastest of these methods, the PCA, to make it robust to the presence of outliers and correct for the bias linked to small datasets. After carefully testing it on mock data, we applied the robust PCA to the 29 MW GCs in our sample.

We report deviation from spherical symmetry for 16 GCs.
Using the $V/\sigma$ diagram, a tool frequently used to link the rotation of elliptical galaxies to their ellipticities, we further analyse the dominant mechanism shaping GCs.
For NGC~3201, NGC~5024, NGC~5286, NGC~5904, NGC~5986, NGC~6205, NGC~6341, and NGC~7078, their location in the $V/\sigma$ diagram, combined with a good alignment between the semi-minor axis and the rotation axis, leads us to conclude that rotation is the main driver of the flattening in these seven GCs.

NGC~104 and NGC~5904 also exhibit a good alignment between their semi-minor axis and rotation axis, reinforcing the idea that rotation plays a dominant role in shaping these clusters, even though their flattening is lower than expected from their rotational signature in the isotropic oblate rotator framework. 

For other clusters such as NGC~6838 and NGC~6656, tidal interactions from the galaxy and/or complex formation history are a plausible explanation for their apparent shape.

We note that our method is both robust and accurate, but it does not account for possible radial variations in ellipticity. Capturing such trends would require extending the method to include radial binning of stars and constructing ellipticity profiles. Although feasible, this extension requires careful consideration to ensure the reliability of the recovered profiles and goes beyond the scope of this study. To ensure that our results and interpretations remain valid using the different radial coverage available, varying between 1.9 $r_h$ and more than 20 $r_h$ for some clusters, we computed the ellipticity using all the stars within 3 half-light radius for 27 clusters with a radial coverage superior to this limit. Apart from a few clusters discussed in Sect.~\ref{subsec:radial_coverage}, which require careful considerations, we find that the interpretation and discussion of our results remain largely valid.

While the isotropic oblate rotator model is a useful way to compare ellipticity and rotation, it was first developed for elliptical galaxies, which usually have enough rotation for the model to apply. Most non-rotating GCs also match the framework quite well, showing low flattening as expected, except NGC~6838, extensively discussed in Sect.~\ref{subsec:group2}. Clusters exhibiting anisotropy also tend to agree within the error-bars, except for NGC~104, whose radial anisotropy likely explains its offset from the relation. This means the framework can still be a good first-order guide for GCs, but deviations could be expected for clusters exhibiting anisotropy.

Obtaining robust flattening measurements and rotation strength of a significant fraction of MW GCs would be interesting to test the isotropic oblate rotator framework and to better constrain the causes of the flattening.
From a modelling point of view, it would be crucial to take into account the kinematic complexity of these systems, including, for example, rotation and pressure anisotropy in dynamical models of GCs. Moreover, a careful exploration of the time evolution of these systems is needed to identify which mechanisms play a significant role in shaping the properties of present-day GCs.

In this first paper of a two-part series, our main goal was to present the method and general results. In a follow-up paper, we will apply this method to study ellipticity differences among multiple stellar populations in GCs, where the available data are limited, hence the need for a method that performs reliably on small datasets (Fréour et al., in prep.).

\begin{acknowledgements}
The authors are very grateful for the referee's careful reading of the manuscript and their valuable comments on the issue of photometric incompleteness and inhomogeneous radial coverage, which significantly improved the quality of this paper. 
The authors would like to thank Francisco Aros for the helpful comments on the manuscript and Paolo Bianchini for the discussions.
EL acknowledges support from the ERC Consolidator Grant funding scheme (project ASTEROCHRONOMETRY, \url{https://www.asterochronometry.eu}, G.A. n. 772293). EP and EL acknowledge funding by the European Union (ERC-2022-AdG, "StarDance: the non-canonical evolution of stars in clusters", Grant Agreement 101093572, PI: E. Pancino). Views and opinions expressed are however those of the author(s) only and do not necessarily reflect those of the European Union or the European Research Council. Neither the European Union nor the granting authority can be held responsible for them. This research has made use of NASA’s Astrophysics Data System (ADS) and of the SIMBAD database operated at the
Strasbourg astronomical Data Center (CDS), Strasbourg (France).\\
\end{acknowledgements}

\section*{Data availability}
Table 1 will be made available through the VizieR database upon acceptance of the manuscript. The Python codes used to obtain the results presented in this article are available upon request.

\bibliographystyle{aa} 
\bibliography{aa_56820-25} 

\begin{appendix}
\onecolumn
\section{Bias correction}
\label{appendixA}

Our aim is to construct a bias correction function using a set of mock data, to correct for the systematic biases linked to small datasets ($N \simeq 100$) and to the overestimation of the ellipticity for clusters close to spherical.
To do this, we proceeded as follows. 
We fixed the angle to $\Phi_{\rm true} = 45^{\circ}$ because, contrary to the ellipticity, changing its value does not impact the results. We generated mock data of GCs with true ellipticity varying from $e_{\rm true} = 0.01$ to $e_{\rm true}=0.2$ with steps of 0.01, and a total number of stars $N$ going from $N = 40$ to $N=1500$ in steps of 20. For each pair of parameters ($e_{\rm true}$,$N$), we created 500 different GCs using a random distribution of stars following a \citet{1962King} profile. We fix the core radius of the clusters to the median value of our sample $r_c = 0.6$ arcmin, taken from \citet{1996Harris}. The truncation radius is also fixed ($r_t = 17.5$ arcmin) and computed from the median value of the concentration of the clusters, also taken from \citet{1996Harris}.
Using the robust PCA algorithm, we computed the ellipticity $e_{\rm rec}$ as the mean value of the ellipticity of the 500 realizations, and the associated error by taking the standard deviation. 
The problem is now a classical regression problem where we seek to find the true ellipticity (target) of an elliptical distribution of stars given a total number of stars and the (biased) recovered ellipticity (features). Several methods can be used to solve such a problem. Among the ones we tried are the polynomial regression (where the relationship between the input variables and the output is a $n$-degree polynomial) and the random forest regression (which creates multiple decision trees, each learning slightly different patterns from the data by using random subsets of data and features, and then averages their predictions to produce a final output).
We find that the random forest regression brings a better correction than the polynomial regression. We also fit a regression model to the residuals to make the correction more accurate, using a Gradient Boosting Regressor, better at capturing complex relationships in data.
Figure~\ref{fig:bias_corr}, representing the true ellipticity versus the ellipticity recovered, illustrates the results of this correction. The blue dots indicate the value of the ellipticity recovered by the robust PCA before correction, while the orange circles represent the value of the ellipticity after bias correction.
\begin{figure}[ht]
\centering
\includegraphics[width=10cm]{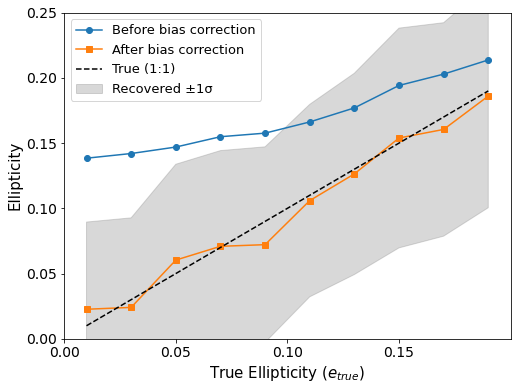}
\caption{This plot shows the ellipticity recovered using the robust PCA method before (blue dots) and after (orange dots) correction for the bias due to a small number of stars. The grey area shows the uncertainties in the recovered ellipticity. The dashed line indicates the true ellipticity. The mock GCs were simulated using 300 stars.}
\label{fig:bias_corr}
\end{figure}

\section{Impact of photometric incompleteness on the results}
\label{sec:appendix_b}
In this article, we used RGB stars only to compute the ellipticity. For most clusters, the full stellar samples are affected by some degree of photometric incompleteness, particularly at the faint end (due to detection limits).

We evaluate the impact of this effect on four representative clusters spanning different sample sizes and ellipticities: NGC~6838 (>100,000 stars in the full sample, $e=0.107 \pm 0.014$), NGC~6254 (~14,000 stars, very round $e=0.025 \pm 0.004$), NGC~4833 (~7,000 stars, round) and NGC~7089 (~4,000 stars, $e = 0.078 \pm 0.006$).
In Fig.~\ref{fig:B_incompleteness}, we plot the distribution of stars as a function of V magnitude and define approximate completeness thresholds for each cluster (in green). We clean the sample by removing stars with V magnitudes above this limit. Then, we run the robust PCA code and compute the ellipticity as described in Sect.~\ref{sec:results}. The results are presented in Table~\ref{tab:2}.

\begin{figure*}
\centering
\includegraphics[width=0.4\textwidth]
{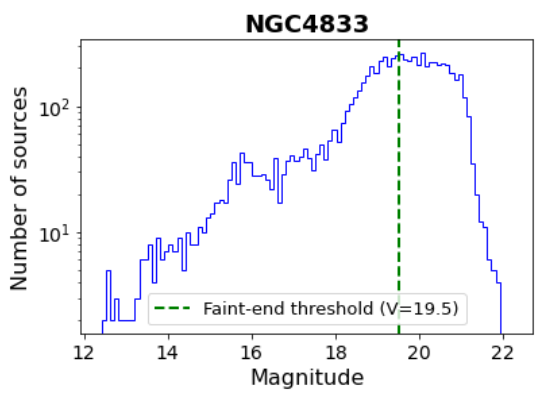}
\includegraphics[width=0.4\textwidth]
{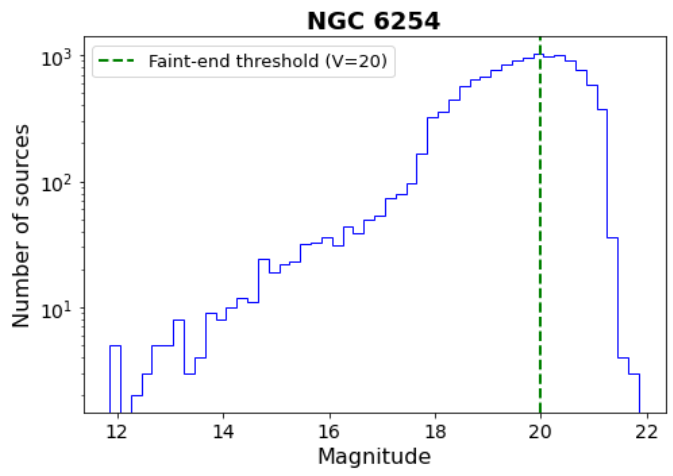}
\includegraphics[width=0.4\textwidth]
{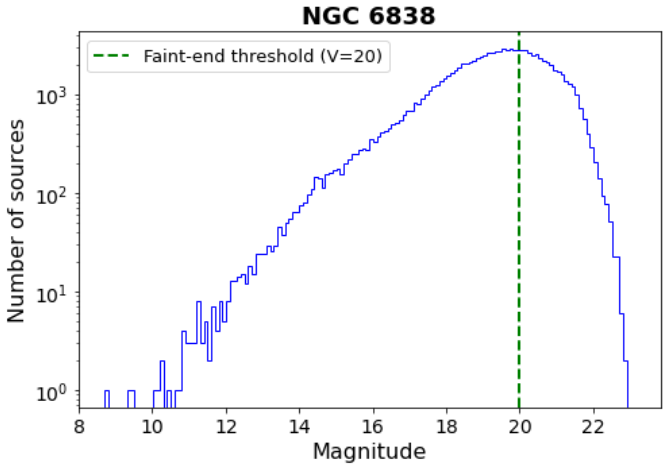}
\includegraphics[width=0.4\textwidth]
{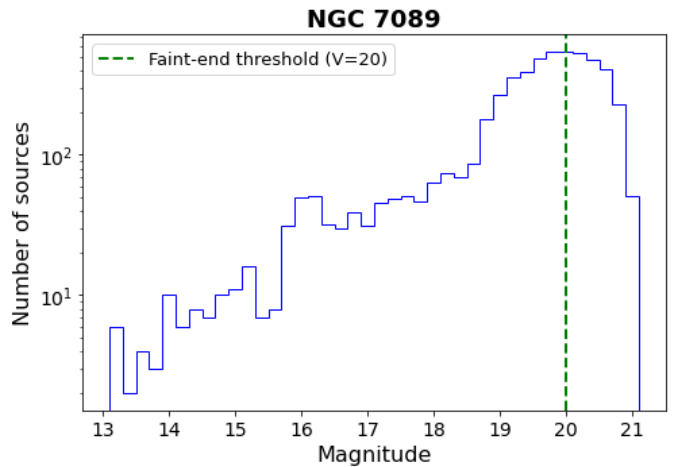}
\caption{The number of sources is plotted per magnitude bins (log scale) for four clusters: NGC~4833 (upper-left), NGC~6254 (upper-right), NGC~6838 (lower-left), and NGC~7089 (lower-right). In each panel, the green dashed lines represent the faint-end threshold cut.}
\label{fig:B_incompleteness}
\end{figure*}

\begin{table*}
 \label{tab:2}
\centering

\caption{Ellipticity measurements for four representative globular clusters using different stellar samples.}
\begin{tabular}{ccccccc}
\hline
Cluster & $N_{all}$& $e_{all}$&$N_{Vcut}$ & $e_{Vcut}$& $N_{RGBs}$&$e_{RGBs}$ \\
\hline
4833  & 7425&$0.045 \pm 0.006$ & 2816& $0.037 \pm 0.011$ & 1055  & $0.14 \pm 0.02$ \\
6254  & 14158&$0.025 \pm 0.004$ & 8111& $0.029 \pm 0.006$ & $1380 $  & $0.033 \pm 0.024$ \\
6838  & 104515 &$0.102 \pm 0.007$ & 66977& $0.092 \pm 0.006$ & 409 & $0.14 \pm 0.025$ \\
7089  & 5821 &$0.078 \pm 0.006$ & 3336& $0.051 \pm 0.025$ & 2709  & $0.043 \pm 0.018$ \\
\hline
\end{tabular}
\tablefoot{The first two columns refer to the total number of stars and ellipticity measured from the full catalogue. $N_{\rm Vcut}$ and $e_{\rm Vcut}$ correspond to the sample restricted by a faint-end magnitude cut to ensure completeness. The last two columns report the number of red giant branch stars and the ellipticity computed using only this subset. All ellipticity values are reported with their $1\sigma$ uncertainties.}
\end{table*}

The table summarises the ellipticity measurements for four representative clusters using three different stellar samples: the full sample ($N_{\rm all}$), the magnitude-limited sample after applying a faint-end cut ($N_{\rm Vcut}$), and the RGB-only sample ($N_{\rm RGBs}$). For NGC~6254, NGC~6838, and NGC~7089, the ellipticity values remain broadly consistent across the datasets, with minor variations.

Interestingly, NGC~4833 shows a notably higher ellipticity when considering only RGB stars ($e_{\rm RGBs} = 0.14 \pm 0.02$) compared to both the full sample ($e_{\rm all} = 0.045 \pm 0.006$) and the magnitude-limited sample ($e_{V{\rm cut}} = 0.037 \pm 0.011$), as shown in Table~\ref{tab:2}. Given that the RGB sample is well-defined and the ellipticity measurement is robust against small-number statistics, this difference likely reflects other factors.
This discrepancy could stem from crowding or radial-dependent incompleteness in the central regions, which are not fully accounted for by the faint-end magnitude cut.

\thispagestyle{empty}
\clearpage

\section{Table of ellipticity results}

\begin{table*}[ht]
\centering
\scriptsize
\caption{This table compares our ellipticity measurements with those provided in \citet{1987White}, \citet{Chen_2010}, and \citet{2024Cruz_reyes}.}
\begin{tabular}{cccccccccccccc}
\hline
Cluster & $N_{\rm tot}$ & $r_h$ & $R_{max}/r_h$ & $e \pm \sigma_e$ & $e_{3r_h} \pm \sigma_{3r_h}$ & $e_{\rm WS87}$ & $e_{\rm CC10}$ & $e_{CR24}$ & $\phi \pm \sigma_{\phi}$ & $\theta_0$ & $i$ \\
\hline
104   & 5484 & 3.23 & 5.1  & $0.076 \pm 0.012$   & $0.038 \pm 0.009$   & $0.09 \pm 0.01$ & $0.16 \pm 0.02$ & $0.059 \pm 0.004$ & $133.7 \pm 3.8$ & $44 \pm 4$ & $33 \pm 2$ \\
288   & 676  & 2.23 & 4.6  & $0.013 \pm 0.023$   & $0.010 \pm 0.022$   & -              & $0.06 \pm 0.01$ & $0.021 \pm 0.010$ & $55.2 \pm 7.9$ & $96 \pm 49$ & $36 \pm 20$ \\
1261  & 1209 & 0.68 & 19.1 & $0.086 \pm 0.012$   & $0.012 \pm 0.018$   & $0.07 \pm 0.00$ & $0.23 \pm 0.07$ & $0.042 \pm 0.018$ & $102.2 \pm 2.9$ & $18 \pm 20$ & $66 \pm 25$ \\
1851  & 1595 & 0.51 & 23.6 & $0.012 \pm 0.011$   & $0.006 \pm 0.024$   & $0.07 \pm 0.00$ & $0.23 \pm 0.07$ & $0.042 \pm 0.018$ & $11.6 \pm 8.2$ & $90 \pm 8$ & $65 \pm 8$ \\
2808  & 4892 & 0.80 & 21.8 & $0.13 \pm 0.007$    & $0.118 \pm 0.010$   & $0.07 \pm 0.00$ & $0.23 \pm 0.07$ & $0.042 \pm 0.018$ & $132.8 \pm 1.8$ & $51 \pm 6$ & $88 \pm 3$ \\
3201  & 716  & 3.10 & 3.55 & $0.094 \pm 0.02$    & $0.125 \pm 0.020$   & $0.12 \pm 0.00$ & $0.20 \pm 0.01$ & $0.065 \pm 0.007$ & $110.0 \pm 5.5$ & $26 \pm 9$ & $64 \pm 5$ \\
4590  & 556  & 1.51 & 5.7  & $0.052 \pm 0.02$    & $0.016 \pm 0.027$   & $0.05 \pm 0.01$ & $0.16 \pm 0.05$ & $0.062 \pm 0.015$ & $170.3 \pm 23.4$ & $176 \pm 31$ & $82 \pm 21$ \\
4833  & 1055 & 2.41 & 3.3  & $0.14 \pm 0.02$     & $0.137 \pm 0.020$   & $0.07 \pm 0.01$ & $0.11 \pm 0.02$ & $0.022 \pm 0.009$ & $15.7 \pm 3.9$ & $124 \pm 83$ & $61 \pm 26$ \\
5024  & 2091 & 1.31 & 9.7  & $0.051 \pm 0.014$   & $0.063 \pm 0.012$   & $0.01 \pm 0.00$ & $0.14 \pm 0.04$ & $0.039 \pm 0.014$ & $138.2 \pm 7.6$ & $119 \pm 32$ & $33 \pm 14$ \\
5053  & 202  & 2.61 & 3.3  & $0.183 \pm 0.029$   & $0.183 \pm 0.028$   & $0.21 \pm 0.01$ & $0.21 \pm 0.03$ & $0.137 \pm 0.023$ & $139.2 \pm 6.6$ & - & - \\
5272  & 2208 & 2.31 & 7.1  & $0.06 \pm 0.015$    & $0.075 \pm 0.015$   & $0.04 \pm 0.00$ & $0.06 \pm 0.01$ & $0.025 \pm 0.008$ & $144.4 \pm 5.8$ & $99 \pm 9$ & $32 \pm 8$ \\
5286  & 2938 & 0.73 & 9.6  & $0.096 \pm 0.009$   & $0.106 \pm 0.008$   & $0.12 \pm 0.01$ & $0.17 \pm 0.01$ & $0.047 \pm 0.013$ & $118.8 \pm 4.5$ & $6 \pm 9$ & $80 \pm 7$ \\
5904  & 1843 & 1.77 & 13.0 & $0.15 \pm 0.02$     & $0.153 \pm 0.017$   & $0.14 \pm 0.00$ & $0.06 \pm 0.02$ & $0.057 \pm 0.008$ & $2.3 \pm 5.2$ & $71 \pm 16$ & $30 \pm 2$ \\
5986  & 2120 & 0.98 & 5.6  & $0.088 \pm 0.014$   & $0.125 \pm 0.017$   & $0.06 \pm 0.00$ & $0.26 \pm 0.11$ & $0.040 \pm 0.013$ & $142.1 \pm 4.8$ & $77 \pm 53$ & $54 \pm 18$ \\
6101  & 612  & 1.05 & 9.2  & $0.019 \pm 0.025$   & $0.000 \pm 0.033$   & $0.05 \pm 0.00$ & $0.06 \pm 0.01$ & $0.066 \pm 0.013$ & $27.0 \pm 12.4$ & $64 \pm 28$ & $63 \pm 15$ \\
6121  & 551  & 4.33 & 1.9  & $0.037 \pm 0.02$    & -                   & $0.00 \pm 0.01$ & $0.07 \pm 0.02$ & $0.018 \pm 0.007$ & $144.6 \pm 38.2$ & $50 \pm 15$ & $87 \pm 12$ \\
6205  & 2181 & 1.69 & 9.6  & $0.10 \pm 0.011$    & $0.091 \pm 0.020$   & $0.11 \pm 0.00$ & $0.12 \pm 0.02$ & $0.050 \pm 0.007$ & $19.1 \pm 4.4$ & $107 \pm 10$ & $16 \pm 3$ \\
6218  & 742  & 1.77 & 5.8  & $0.009 \pm 0.019$   & $0.007 \pm 0.054$   & $0.04 \pm 0.00$ & $0.08 \pm 0.01$ & $0.017 \pm 0.008$ & $10.6 \pm 10.0$ & $173 \pm 75$ & $22 \pm 12$ \\
6254  & 1380 & 1.95 & 8.6  & $0.033 \pm 0.024$   & $0.014 \pm 0.021$   & $0.00 \pm 0.00$ & $0.11 \pm 0.04$ & $0.025 \pm 0.008$ & $21.2 \pm 10.5$ & $14 \pm 80$ & $26 \pm 16$ \\
6341  & 1212 & 1.02 & 19.2 & $0.11 \pm 0.017$    & $0.012 \pm 0.017$   & $0.10 \pm 0.00$ & $0.11 \pm 0.03$ & $0.042 \pm 0.010$ & $136.5 \pm 8.2$ & $44 \pm 29$ & $82 \pm 10$ \\
6366  & 606  & 2.92 & 5.6  & $0.030 \pm 0.018$   & $0.000 \pm 0.026$   & $0.16 \pm 0.01$ & $0.09 \pm 0.04$ & $0.030 \pm 0.011$ & $62.0 \pm 46.8$ & $44 \pm 84$ & $51 \pm 20$ \\
6656  & 968  & 3.36 & 2.2  & $0.012 \pm 0.022$   & -                   & $0.14 \pm 0.01$ & $0.12 \pm 0.01$ & $0.056 \pm 0.005$ & $106.2 \pm 15.9$ & $9 \pm 7$ & $18 \pm 3$ \\
6809  & 809  & 2.83 & 4.7  & $0.01 \pm 0.022$    & $0.010 \pm 0.025$   & $0.02 \pm 0.00$ & $0.11 \pm 0.01$ & $0.015 \pm 0.006$ & $97.5 \pm 18.3$ & $41 \pm 72$ & $74 \pm 14$ \\
6838  & 409  & 1.67 & 4.3  & $0.14 \pm 0.025$    & $0.184 \pm 0.028$   & $0.00 \pm 0.01$ & $0.32 \pm 0.02$ & $0.016 \pm 0.008$ & $17.1 \pm 3.7$ & $20 \pm 97$ & $34 \pm 30$ \\
6934  & 820  & 0.69 & 8.7  & $0.008 \pm 0.015$   & $0.010 \pm 0.029$   & $0.01 \pm 0.00$ & $0.06 \pm 0.02$ & $0.031 \pm 0.016$ & $55.4 \pm 8.4$ & - & - \\
6981  & 542  & 0.93 & 7.1  & $0.016 \pm 0.030$   & $0.015 \pm 0.027$   & $0.02 \pm 0.00$ & $0.14 \pm 0.09$ & $0.034 \pm 0.018$ & $157.0 \pm 12.8$ & - & - \\
7078  & 2372 & 1.00 & 16.8 & $0.062 \pm 0.018$   & $0.080 \pm 0.015$   & $0.07 \pm 0.00$ & $0.23 \pm 0.07$ & $0.042 \pm 0.018$ & $119.3 \pm 5.5$ & $37 \pm 10$ & $62 \pm 5$ \\
7089  & 2709 & 1.06 & 11.1 & $0.043 \pm 0.018$   & $0.032 \pm 0.015$   & $0.11 \pm 0.00$ & $0.14 \pm 0.02$ & $0.027 \pm 0.011$ & $54.0 \pm 7.0$ & $119 \pm 7$ & $18 \pm 5$ \\
7099  & 521  & 1.03 & 8.6  & $0.015 \pm 0.040$   & $0.000 \pm 0.026$   & $0.01 \pm 0.01$ & $0.10 \pm 0.01$ & $0.032 \pm 0.012$ & $34.9 \pm 22.0$ & $86 \pm 58$ & $18 \pm 13$ \\
\hline
\end{tabular}
\tablefoot{The NGC number is given in the first column. In the second column, we provide the total number of stars available in the complete RGB sample. $r_h$ is the half-light radius in arcmin taken from \cite{1996Harris}. $R_{max}/r_h$ is the ratio of the maximum distance to the cluster's centre and the half-light radius. The next five columns present the ellipticities from our analysis using all stars ($e$), restricted to within three half-light radii ($e{3r_h}$, as well as the ellipticities reported by WS87, CC10, and CR24. We report on the position angle of the major axis of the ellipse in the ninth column (in degrees). The last columns correspond to the position angle of the rotation axis and the inclination angle of the cluster, both in degrees, after applying the modifications described in Sect.~\ref{sec:results} to the values reported in Table A.1 from \citet{2024Leitinger}. We find a median flattening of 0.052.}
\label{tab:cluster_results}
\end{table*}

\end{appendix}

%

%

\end{document}